\documentclass{article}
\usepackage{spconf,amsmath,graphicx,cite,amssymb,supertabular,multirow,bbding,xcolor}

\title{Multi-Processor Approximate Message Passing Using Lossy Compression}

\name{Puxiao Han,$^{v}$ Junan Zhu,$^{n}$ Ruixin Niu,$^{v}$
and Dror Baron$^{n}$\thanks{The work of P. Han and R. Niu was supported in 
part by the VCU Presidential Research
Quest Fund. The work of J. Zhu and D. Baron was supported in part by the National Science Foundation under Grant CCF-1217749 and the U.S. Army Research Office under  Contract W911NF-14-1-0314.}}
\address{{\small $v$: Virginia Commonwealth University, Dept. Electrical and Computer Engineering}\\
	{\small Richmond, VA 23284, U.S.A. Email: \{hanp, rniu\}@vcu.edu}\\
	{\small $n$: North Carolina State University, Dept. Electrical and Computer Engineering} \\
	{\small Raleigh, NC 27695, U.S.A. Email: \{jzhu9, barondror\}@ncsu.edu} \\
	}
\begin{document}
%\ninept
%
\maketitle
\begin{abstract}
\vspace*{-1mm}
In this paper, a communication-efficient multi-processor compressed sensing framework based on the approximate message passing algorithm is proposed. We perform lossy compression on the data being communicated between processors, resulting in a reduction in communication costs with a minor degradation in recovery quality. In the proposed framework, a new state evolution formulation takes the quantization error into account, and analytically determines the coding rate required in each iteration. Two approaches for allocating the coding rate, an online back-tracking heuristic and an optimal allocation scheme based on dynamic programming, provide significant reductions in communication costs.
\end{abstract}
\section{Introduction}
\label{sec:intro}

Compressed sensing (CS) \cite{donoho2006compressed,candes2006stable} has 
numerous applications in various areas of signal processing. Due to the curse of dimensionality, it can be demanding to perform CS on a single processor. 
Furthermore, clusters comprised of many processors have the
potential to accelerate computation.  Hence, multi-processor CS (MP-CS) has become of recent interest~\cite{DBP,DIHT,patterson2014distributed}. 

%\vspace{-2mm}
We consider MP-CS systems comprised of two parts: 
({\em i}) local computation (LC) is performed at each processor, and 
({\em ii}) global computation (GC) obtains an estimate of the unknown 
signal after processors exchange the results of LC.  In our previous work \cite{han2014distributed}, we developed an MP-CS framework based 
on the approximate message passing (AMP)
algorithm~\cite{AMP}, in which a GC approach performs
AMP in an MP system, providing the same recovery result as centralized AMP. 
We chose AMP, because it is analytically tractable due to the 
state evolution (SE) \cite{TUNE,Dynamic} formalism, and can be extended to Bayesian CS \cite{Dynamic,tan2015compressive}, matrix completion \cite{parker2013bilinear}, and non-negative principal component analysis (PCA) \cite{montanari2014non}. 
%\vspace{-2mm}

Compared with many results on distributed computation, optimization, and network topology \cite{DBP,mota2013d} in MP-CS, a modest subset of the literature considers the communication costs of the GC step \cite{DIHT, patterson2014distributed, han2015modified,han2015communication}. In this paper, we still consider multi-processor AMP (MP-AMP), and focus more on the communication costs. In contrast to our prior work
\cite{han2014distributed}, we are willing to accept a minor decrease in 
recovery quality
while providing significant and often 
dramatic communications savings. Such results are especially
well-suited to clusters where communication between servers is costly.
To achieve the reduction in communication costs, 
we use lossy compression to reduce the inter-processor communication costs, 
and provide a modified SE formulation that accounts for quantization error. Two approaches for allocating the coding rate, an online back-tracking heuristic and an optimal allocation scheme based on dynamic programming, provide significant reductions in communication costs. Furthermore, we consider Bayesian AMP, which achieves better recovery accuracy than non-Bayesian AMP~\cite{AMP} by assuming that the unknown signal follows a known prior distribution.
%\vspace{-0.3mm}

In the following, bold capital and bold lower-case letters are used to denote matrices and vectors respectively, capital letters without bold typically refer to dimensionality or random variables, and $[\cdot]^T$ denotes vector or matrix transposition. 
The $\ell_2$ norm of a vector is denoted by
$\|\cdot\|$, ${\cal N}(\mu,\sigma^2)$ is a Gaussian 
distribution with mean $\mu$ and variance $\sigma^2$, and ${\cal U}\left[a,b\right]$ is a continuous uniform 
distribution within $\left[a,b\right]$.
%\vspace{-0.3mm}

\section{The Centralized AMP Algorithm}
%\subsection{CS recovery problems}
Approximate message passing (AMP)~\cite{AMP} 
is a statistical algorithm derived from the theory of probabilistic graphical models \cite{koller2009probabilistic}. Given noisy measurements $\mathbf{y}=\mathbf{As}_0+\mathbf{e}$ of the unknown signal $\mathbf{s}_0\in\mathbb{R}^N$, where elements in $\mathbf{s}_0$ are 
independent and identically distributed
(i.i.d.) realizations of a scalar random variable $S_0 \sim p_{S_0}$, $\mathbf{A}\in\mathbb{R}^{M\times N}$ is the sensing matrix with entries $\sim$ i.i.d. ${\cal N}(0,1/M)$, and $\mathbf{e}\in \mathbb{R}^M$ is additive measurement noise, which is i.i.d. ${\cal N}(0,\sigma^2_e)$, AMP iteratively recovers $\mathbf{s}_0$, starting from an initial estimate $\mathbf{x}_0 = 0$ and residual $\mathbf{z}_0=\mathbf{y}$:
\vspace{-2mm}
\begin{equation}
\label{eqn:AMP_ft}
\mathbf{f}_{t}= \mathbf{x}_t+\mathbf{A}^{T} \mathbf{z}_t,
\end{equation}
\vspace{-5mm}
\begin{equation}
\label{eqn:AMP_new_x}
\mathbf{x}_{t+1}= \eta_t(\mathbf{f}_t),
\end{equation}
\vspace{-5mm}
\begin{equation}
\label{eqn:AMP_new_z}
\mathbf{z}_{t+1}= \mathbf{y} - \mathbf{Ax}_{t+1} + (N/M) \overline{\eta'_t(\mathbf{f}_t)} \mathbf{z}_{t},
\end{equation}
where $t$ is the iteration number,
the bar above the vector in \eqref{eqn:AMP_new_z} denotes its 
empirical average, $\eta_t$ is known as the denoising function or denoiser, and $\eta'_t$ denotes its derivative.

According to Bayati and Montanari \cite{Dynamic}, as $N\rightarrow \infty$ and $M/N=\kappa>0$, the elements of $\mathbf{f}_t$ in \eqref{eqn:AMP_ft} follow i.i.d. $F_t= S_0 + \sigma_t Z$, where $Z\sim {\cal N}(0,1)$ and the sequence $\{\sigma^2_t\}$ satisfies
\vspace{-3mm}
\begin{equation}
	\label{eqn:SE}
	\begin{split}
		& \sigma^2_{t+1} =\sigma^2_e+(1/\kappa)\mathbb{E}\left[\eta_t(S_0+\sigma_{t} Z)-S_0\right]^2\\
		&=\sigma^2_e+(1/\kappa)\mathbb{E}\|\mathbf{x}_{t+1}-\mathbf{s}_0\|^2/N.
	\end{split}
\end{equation}
Note that $\sigma^2_0 =\sigma^2_e+(1/\kappa)\mathbb{E}\left[S_0\right]^2$; equation \eqref{eqn:SE} is known as state evolution (SE), and the optimal denoiser for mean square error (MSE) is the conditional mean \cite{Dynamic,tan2014compressive}:
\vspace{-1mm}
\begin{equation}
\label{eqn:denoiser}
\eta_t(F_t) = \mathbb{E} \left[S_0 \left| S_0 +\sigma_t Z = F_t\right.\right].
\end{equation}

In this paper, we assume that $S_0$ follows the Bernoulli Gaussian distribution:
\vspace{-2mm}
\begin{equation}
\label{eqn:pS0}
p_{S_0}(s) = \epsilon {\cal N}(s;\mu_s,\sigma^2_s) + (1-\epsilon) \delta(s),
\end{equation}
where $\delta(s)$ denotes the Dirac delta function, and $S_0$ typically has mean $\mu_s=0$. The denoiser is easily derived using \eqref{eqn:denoiser}. 

As a measure of the measurement noise level and recovery accuracy, we define the signal-to-noise-ratio (SNR) as
\vspace{-3mm}
\begin{equation}
	\begin{split}
		& \text{SNR}=10\log_{10}\left({\mathbb{E}\left[\|\mathbf{As}_0\|^2\right]}/{\mathbb{E}\left[\|\mathbf{e}\|^2\right]}\right) \\
		& \approx 10\log_{10}\left({\mathbb{E}\left[\|\mathbf{s}_0\|^2\right]}/{\mathbb{E}\left[\|\mathbf{e}\|^2\right]}\right)=10\log_{10}\left( {\rho}/{\sigma^2_e}\right),
	\end{split}
	\nonumber
	\vspace{-5mm}
\end{equation}
where $\rho=\epsilon/\kappa$, and the signal-to-distortion-ratio (SDR) at iteration $t$ as
\vspace{-3mm}
\begin{equation}
	\label{eqn:SNR_tval}
	\text{SDR}(t)=10\log_{10}\left({\mathbb{E}\left[\|\mathbf{s}_0\|^2\right]}/{\mathbb{E}\left[\|\mathbf{x}_t-\mathbf{s}_0\|^2\right]}\right). \nonumber
	\vspace{-2mm}
\end{equation}
Using the SE equation in \eqref{eqn:SE}, we have
\vspace{-3mm}
\begin{equation}
	\text{SDR}(t)=10\log_{10} \left[{\rho}/\left({\sigma^{2}_{t}-\sigma^2_e}\right)\right]. \nonumber
	\vspace{-2mm}
\end{equation}
Note that the Bernoulli Gaussian assumption in this paper is only for illustration, and our work is easily extended to other prior distributions $p_{S_0}$.

\section{Multi-Processor AMP Framework}
\subsection{Communication in Multi-Processor AMP}
Consider a system with $P$ processors and one fusion center. Each processor $p\in\{1,\cdots,P\}$ takes $M/P$ rows of $\mathbf{A}$, namely $\mathbf{A}^p$, and obtains $\mathbf{y}^p=\mathbf{A}^{p} \mathbf{s}_0+\mathbf{e}^p$. The procedures in \eqref{eqn:AMP_ft} --- \eqref{eqn:AMP_new_z} can then be rewritten in a distributed manner:

{\emph {Local Computation (LC) performed by each processor $p$}}:
\vspace{-2mm}
\begin{equation}
\label{eqn:DAMP_z}
\mathbf{z}^p_{t}=\mathbf{y}^p-\mathbf{A}^p \mathbf{x}_{t}+(1/\kappa) \overline{\eta'_t(\mathbf{f}_{t-1})} \mathbf{z}^p_{t-1},\nonumber
\end{equation}
\vspace{-5mm}
\begin{equation}
	\label{eqn:DAMP_W}
	\mathbf{f}^p_t=\mathbf{x}_t/P + (\mathbf{A}^{p})^{T}\mathbf{z}^p_t.\nonumber
\end{equation}
%By expressing the system in this way, a distributed framework is presented like this: . We adopt this framework and 
%It is easy to show that AMP can be run in a distributed manner:
{\emph {Global Computation (GC) performed by the fusion center}}:
\vspace{-3mm}
\begin{equation}
	\label{eqn:DAMP_ft}
	\mathbf{f}_t= \sum_{p=1}^P \mathbf{f}^p_t,\text{   }\overline{\eta'_t(\mathbf{f}_{t})},{\text{   and   }}\mathbf{x}_{t+1}= \eta_t\left(\mathbf{f}_t\right).\nonumber
	\vspace{-3mm}
\end{equation}

It can be seen that in the GC step of MP-AMP, each processor $p$ sends $\mathbf{f}^p_t$ to the fusion center, and the fusion center sums them to obtain $\mathbf{f}_t$ and $\mathbf{x}_{t+1}$, and sends $\mathbf{x}_{t+1}$ to each processor.{\footnote{In order to calculate each $\mathbf{z}^p_{t+1}$, the fusion center also needs to send $\overline{\eta'_t(\mathbf{f}_{t})}$ to all the processors. This is a scalar, and the corresponding communication cost is negligible compared with that of transmitting a vector.}} Our goal in this paper is to reduce these communication costs while barely impacting recovery performance.

Suppose that all the elements in $\mathbf{f}^p_t$ are computed as $32$-bit single-precision floating-point numbers.  Because SE is robust to small perturbations \cite{AMP,TUNE}, we can compress $\mathbf{f}^p_t$ lossily up to some reasonable distortion level, 
and send the compressed output to the fusion center. To ensure that this error is 
indeed a ``small perturbation,"  we require the error to be additive and, 
if possible, white and Gaussian, so that we can analyze the relationship between 
the error and coding rate.

\subsection{Lossy Compression of $\mathbf{f}^p_t$}
\label{sec:qft}
Before we propose specific lossy compression approaches, we 
describe an important property of MP-AMP. In addition to the well-known Gaussianity of the vector $\mathbf{f}_t-\mathbf{s}_0$ \cite{Dynamic}, numerical results show that elements 
of $\mathbf{f}^p_t-(1/P)\mathbf{s}_0$ are also i.i.d. Gaussian with mean $0$ and variance $\sigma^2_t/P$. Furthermore, $\mathbf{f}^p_t-(1/P)\mathbf{s}_0$ and $\mathbf{f}^q_t-(1/P)\mathbf{s}_0$ are independent for different processors $p$ and $q$. In light of this property, $\mathbf{f}^p_t$ can be described as a scalar channel:
\vspace{-3mm}
\begin{equation}
\label{eqn:fpt_scalar_channel}
F^p_t = S_0/P + ({\sigma_t}/{\sqrt{P}}) Z_p,\text{     where     }Z_p\sim {\cal N}(0,1).\nonumber
\vspace{-2mm}
\end{equation}
%\subsection{The estimator of $\sigma^2_t$ in DAMP}
For the Bernoulli Gaussian distribution \eqref{eqn:pS0},
\vspace{-3mm}
\begin{equation}
\label{eqn:fpt_distribution}
F^p_t \sim \epsilon {\cal N}\left(\mu_s/P,(\sigma^2_s+P\sigma^2_t)/P^2\right) + (1-\epsilon) {\cal N}\left(0,\sigma^2_t/P\right).\nonumber
\vspace{-2mm}
\end{equation}

\noindent{\bf Scalar Quantization:} Next, we propose a uniform quantizer with entropy coding, also known as entropy coded scalar quantization (ECSQ) \cite{gersho2012vector}. 

Let $\Psi(u)$ denote the characteristic function of $F^p_t$, it can be shown that
\vspace{-3mm}
\begin{equation}
	\begin{split}
		& |\Psi(u)|\leq \epsilon\exp\left[-0.5\left(\sigma^2_s+P\sigma^2_t\right)u^2/P^2\right]\\
		& +(1-\epsilon)\exp{\left(-0.5\sigma^2_t u^2/P\right)}\leq \exp{\left(-0.5\sigma^2_t u^2/P\right)}\nonumber
	\end{split}
	\vspace{-6mm}
\end{equation} 
is nearly band-limited. Due to this property, it is possible to develop a uniform quantizer of $\mathbf{f}^p_t\sim$ i.i.d. $F^p_t$, where the quantization error $\mathbf{v}^p_t$ is approximately statistically equivalent to a uniformly distributed noise $V^p_t \sim {\cal U}\left[-0.5\Delta_Q,0.5\Delta_Q\right]$ uncorrelated to $F^p_t$. Actually, a quantization bin size $\Delta_Q\leq 2\sigma_t/\sqrt{P}$ will suffice for validation of $\mathbf{v}^p_t\sim$ i.i.d. $V^p_t$ ~\cite{widrow2008quantization}.

The fusion center will receive the quantized data
$\widetilde{\mathbf{f}}^p_t\sim \text{i.i.d. } \widetilde{F}^p_t$, and calculate $\widetilde{\mathbf{f}}_t = \sum_{p=1}^P\widetilde{\mathbf{f}}^p_t\sim \text{i.i.d. } \widetilde{F}_t$, where
\vspace{-3mm}
\begin{equation}
\label{eqn:qf_scalar_channel}
\widetilde{F}_t = \sum_{p=1}^{P} \widetilde{F}^p_t = F_t + V_t, \text{  and }
V_t = \sum_{p=1}^{P} V^p_t.
\vspace{-2mm}
\end{equation}
Applying the central limit theorem, $V_t$ approximately follows ${\cal N}(0,P\sigma^2_Q)$ for large $P$, where $\sigma^2_Q=\Delta^2_Q/12$.

\noindent{\bf Entropy Coding and Optimum Bit Rate:} Let $p_i$ be the probability that $F^p_t$ falls into the $i$-th quantization bin. The entropy of quantized $F^p_t$, $\widetilde{F}^p_t$, is $H_Q = -\sum_{i} p_i \log_2 {(p_i)}$\cite{cover2012elements}, that is, the sensors need $H_Q$ bits on average to present each element in $\widetilde{\mathbf{f}}^p_t$ to the fusion center, which is achievable through entropy coding \cite{cover2012elements}.

In rate distortion (RD) theory~\cite{cover2012elements}, we are given a length-$n$ random sequence $Y_n=\{Y_{n,i}\}_{i=1}^n\sim$ i.i.d. $Y$, 
and our goal is to identify a reconstruction sequence $\widehat{Y}_n=\{\widehat{Y}_{n,i}\}_{i=1}^n$ that can be encoded at low rate while the
distortion $d(Y_n,\widehat{Y}_n)=\frac{1}{n}\sum_{i} d(Y_{n,i},\widehat{Y}_{n,i})$ (e.g., squared error distortion) between the input and reconstruction sequences is small.
RD theory has characterized the fundamental best-possible trade-off between the distortion $D=d(Y_n,\widehat{Y}_n)$ and coding rate $R(D)$,
which is called the rate distortion function. The RD function $R(D)$ can be computed numerically (cf.~Blahut~\cite{blahut1972computation} and Arimoto~\cite{arimoto1972algorithm}). 
For the uniform quantizer that yields a quantization MSE of $\sigma^2_Q$ with a coding rate $H_Q$ bits per element, the RD function will give a bit rate $R(D=\sigma^2_Q)<H_Q$, which is achievable through vector quantization \cite{gersho2012vector}.

\noindent{\bf New SE Equation:} For both ECSQ and RD-based vector quantization that lead to a quantization MSE of $\sigma^2_Q$, the fusion center will have $\widetilde{F}_t = S_0 + \sqrt{\sigma^2_t + P\sigma^2_Q} \widetilde{Z}$, where $\widetilde{Z}\sim{\cal N}(0,1)$. The new denoiser and SE equation become
\vspace{-3.5mm}
\begin{equation}
\label{eqn:new_denoiser}
\eta^Q_t(\widetilde{F}_t) = \mathbb{E} \left[S_0 \left| S_0 +\sqrt{\sigma^2_t + P\sigma^2_Q} \widetilde{Z} = \widetilde{F}_t\right.\right]\text{   and}\nonumber
\vspace{-1.5 mm}
\end{equation}
\vspace{-4mm}
\begin{equation}
\label{eqn:new_SE}
\sigma^2_{t+1}\!=\!\sigma^2_e\!+\!(1/\kappa)\mathbb{E}\left[\eta^Q_t\!\left(S_0+\sqrt{\sigma^2_t + P\sigma^2_Q} \widetilde{Z}\!\right)\!-\!S_0\right]^2.
\vspace{-2mm}
\end{equation}

Currently, we only consider compression of $\mathbf{f}^p_t$. When broadcast from the fusion center to the $P$ processors is allowed in the network topology, the communication cost of sending $\mathbf{x}_t$ -- even uncompressed -- is smaller than that of communicating the $P$ vectors $\mathbf{f}^p_t$. We are considering the case where broadcast is not allowed in our ongoing work.
\subsection{Online Back-tracking (BT-MP-AMP)}
\label{sec:BT}
Let $\sigma^2_{t,C}$ and $\sigma^2_{t,D}$ denote the $\sigma^2_t$ obtained by 
centralized AMP \eqref{eqn:SE} and MP-AMP \eqref{eqn:new_SE}, respectively. 
In order to reduce communication while maintaining high fidelity, 
we first constrain $\sigma^2_{t,D}$ so that it will not deviate much from $\sigma^2_{t,C}$, and then determine the minimum coding rate required in each iteration. This can be done through an online back-tracking algorithm, which we name BT-MP-AMP and present below.

In each iteration $t$, before quantizing $\mathbf{f}^p_t$, we first compute $\sigma^2_{t+1,C}$ for the next iteration. Then we find the maximum quantization MSE $\sigma^2_Q$ allowed so that the ratio $\sigma^2_{t+1,D}/\sigma^2_{t+1,C}$ does not exceed some constant, provided that the required bit rate does not exceed some threshold. Based on the obtained $\sigma^2_Q$ we construct the corresponding quantizer.

Note that the SE in \eqref{eqn:new_SE} is only an approximation, and we do not know the true value of $\sigma^2_{t,D}$ in the current iteration. To better predict 
$\sigma^2_{t+1,D}$, we use $\widehat{\sigma}^2_{t,D}=\|\mathbf{z}^p_t\|^2/M$, 
which is a good estimator for $\sigma^2_{t,D}$\cite{TUNE,Dynamic}, to compute $\sigma^2_{t+1,D}$. To obtain $\widehat{\sigma}^2_{t,D}$, each processor $p$ sends the scalar $\|\mathbf{z}^p_t\|^2$ to the fusion center, which then sends the scalar $\hat{\sigma}^2_{t,D}=\sum_{p=1}^{P}\|\mathbf{z}^p_t\|^2/M$ to all the processors. The corresponding communication cost is also negligible compared with that of communicating $\mathbf{f}^p_t$.

\subsection{Dynamic Programming (DP-MP-AMP)}
\label{sec:dp}
While back-tracking is a useful heuristic, it is possible for a given coding budget $R$ per element, total number of AMP iterations $T$, and initial noise level $\sigma^2_0$ in the scalar channel to compute the coding rate allocations among the AMP iterations that minimize the final MSE, $\sigma^2_{T,D}$. 

To do so, note that we can evaluate $\sigma^2_{t,C}$ offline and hence obtain the number of iterations required to reach the steady state, which would be a reasonable choice for $T$. Second, recalling the new SE equation in \eqref{eqn:new_SE}, $\sigma^2_{t,D}$ depends on $\sigma^2_{t-1,D}$ and $\sigma^2_Q$, which is a function of $R_t$, the coding rate allocated in the $t$-th iteration. Therefore, we can rewrite $\sigma^2_{t,D}$ as follows:
\vspace{-1.5mm}
\begin{equation}
\label{eqn:SErewrite}
\begin{split}
& \sigma^2_{t,D} = f_1(\sigma^2_{t-1,D},R_t) = f_2(\sigma^2_{t-2,D},R_{t-1},R_t) \\
& = \cdots = f_t(\sigma^2_0,R_1,\cdots,R_{t-1},R_t),
\end{split}
\vspace{-4mm}
\end{equation}
that is, given $\sigma^2_0$, $\sigma^2_{T,D}$ is only a function of $R_t$ for $t\in\{1,2,\cdots,T\}$. Denoting ${\cal F}_T(R) = \{R_1,\cdots,R_T\geq 0\colon$ $\sum_{t=1}^{T} R_t = R\}$, minimizing $\sigma^2_{T,D}$ for a given $R$ can be formulated as the following optimization problem:
\begin{equation}
\label{eqn:SEopt}
\min\limits_{{\cal F}_T(R)}\sigma^2_{T,D} = \min\limits_{{\cal F}_T(R)}f_T(\sigma^2_0,R_1,\cdots,R_T).
\vspace{-3mm}
\end{equation}
Since $\sigma^2_{t,D}$ is increasing with $\sigma^2_{t-1,D}$, it is easy to verify the following recursive relationship:
\vspace{-2.5mm}
\begin{equation}
\label{eqn:recur}
\!\min\limits_{{\cal F}_T(R)}\!\sigma^2_{T,D}\!=\!\min\limits_{0\leq R_T\leq R}\! f_1\!\left(\!\min\limits_{{\cal F}_{T-1}(R-R_T)}\! \sigma^2_{T-1,D}, \! R_T\right)\!=\!\cdots,\nonumber
\vspace{-1mm}
\end{equation}
which makes the problem solvable through dynamic programming (DP). 

To implement DP, we need to discretize ${\cal F}_T(R)$ into $\{R_1, \cdots, R_T\in\Omega: \sum_{t=1}^T R_t = R\}$, where $\Omega=\{R^{(1)},\cdots,$ $R^{(S)}\}$ with $R^{(s)}=R(s-1)/(S-1)$, $\forall s\in\{1,\cdots,S\}$. In this paper, we set the bit rate resolution $\Delta R = R/(S-1) = 0.1$ bits per element. Then, we create an $S\times T$ array $\mathbf{\Sigma}$, with the element in the $s$-th row ($s\in\{1,\cdots,S\}$) and $t$-th column ($t\in\{1,\cdots,T\}$) denoted as $\sigma^2_D(s,t)$, storing the optimal value of $\sigma^2_{t,D}$ when a total of $R^{(s)}$ bits per element are used in the first $t$ iterations. By definition of $\sigma^2_D(s,t)$, we have
\vspace{-2mm}
\begin{equation}
\label{eqn:dp}
\sigma^2_D(s,t) = \min\limits_{r\in\{1,2,\cdots,s\}} f_1\left(\sigma^2_D(r,t-1), R^{(s-r+1)}\right),
\vspace{-2mm}
\end{equation}
and the first column of elements in $\mathbf{\Sigma}$ is obtained by:
\vspace{-2mm}
\begin{equation}
\label{eqn:first}
\sigma^2_D(s,1) = f_1\left(\sigma^2_0, R^{(s)}\right),\text{   }\forall s\in\{1,2,\cdots,S\}.
\vspace{-2mm}
\end{equation}
After obtaining $\mathbf{\Sigma}$, the optimal value of $\sigma^2_{T,D}$, by definition, is $\sigma^2_D(S,T)$. Meanwhile, to obtain the optimal bit allocation strategy, we need another $S\times T$ array $\mathbf{R}$ to store the optimal bit rate $R_{DP}(s,t)$ that is allocated at iteration $t$ when a total of $R^{(s)}$ bits per element are used in the first $t$ iterations. Similar to BT-MP-AMP, we name the proposed MP-AMP approach combined with DP as DP-MP-AMP.
\section{Numerical Results}
\begin{figure*}[!t]
	\centering
	\includegraphics[width=7 in]{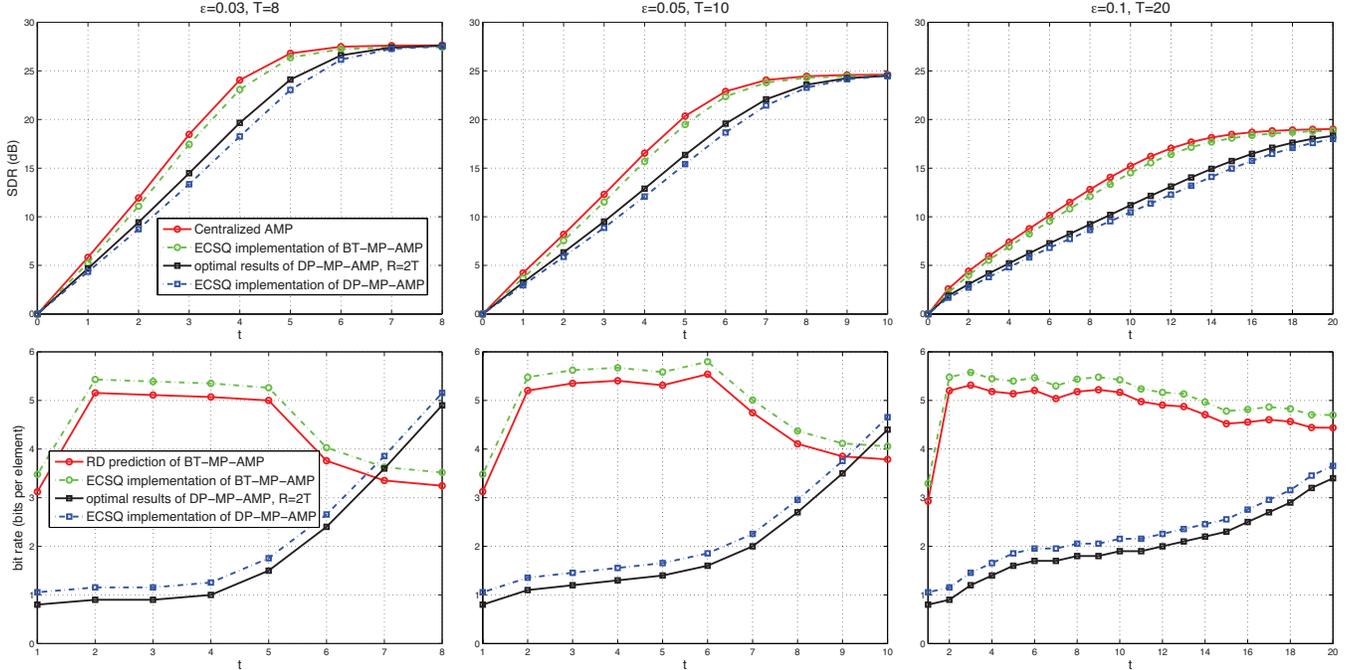}
	% where an .eps filename suffix will be assumed under latex, 
	% and a .pdf suffix will be assumed for pdflatex; or what has been declared
	% via \DeclareGraphicsExtensions.
	\vspace{-0.25in}
	\caption{SDR and bit rates as functions of iteration number $t$. ($N\!=\!10,\!000,M\!=\!3,\!000,\kappa\!=\!0.3,\mu_s\!=\!0,\sigma_s\!=\!1$, $\text{SNR}\!=\!20$ dB.)}
	\vspace{-0.2in}
	\label{fig:SDR}
	
\end{figure*}
\begin{table}[!t]
%	\vspace{-0.1in}
	\caption{ Total bits per element of MP-AMP }
	\small
	\begin{center}
		\label{tb:Rtotal}
		\begin{tabular}{ | l | c | c | c |}
			\hline
			$\epsilon$& $0.03$ & $0.05$ & $0.10$\\ \hline
			$T$& $8$ & $10$ & $20$\\ \hline
			BT-MP-AMP (RD prediction) &$33.82$&$46.43$&$96.16$ \\ \hline 
			BT-MP-AMP (ECSQ simulation) &$36.09$&$49.19$&$101.50$ \\ \hline 
			DP-MP-AMP (RD prediction) &$16$&$20$&$40$ \\ \hline 
			DP-MP-AMP (ECSQ simulation) &$18.04$&$22.55$&$45.10$ \\ \hline 
		\end{tabular}
	\end{center}
	\vspace{-0.2 in}
\end{table}
We evaluate BT-MP-AMP and DP-MP-AMP in an MP system with $P=30$ processors at SNR$=20$ dB, where we set $N = 10,\!000$, $M=3,\!000$, i.e., $\kappa=0.3$, and generate Bernoulli-Gaussian sequences $\mathbf{s}_0$ with $\epsilon\in\{0.03,0.05,0.1\}$, $\mu_s=0$, and $\sigma_s=1$.  

We first evaluate the SE equation \eqref{eqn:SE} of centralized AMP for the three sparsity levels. As shown in Fig. \ref{fig:SDR}, they reach the steady state after $T=8$, $10$, and $20$ iterations respectively. Then, we run BT-MP-AMP and DP-MP-AMP, where for the latter the total rates are $R=2T$ bits per element and the RD-function models the relation between $R_t$ and $\sigma^2_Q$. 

According to RD theory, in the high rate limit, we should expect a gap of roughly $0.255$ bits per element between the entropy and RD function for a given distortion level \cite{gersho2012vector}. Therefore, in an implementation of DP-MP-AMP where we apply ECSQ, we add $0.255$ bits per element to the results in each iteration obtained by DP. Note that the two solid curves in the top three panels are obtained through offline calculation and optimization, and the two dash-dotted curves are obtained through AMP simulations. 

As shown in Fig. \ref{fig:SDR}, BT-MP-AMP uses fewer than $6$ bits per element in each iteration, more than $80\%$ communication savings compared with $32$-bit single-precision floating-point transmission, while achieving almost the same SDR's as in centralized AMP. On the other hand, there are clear gaps between the SDR's of DP-MP-AMP and centralized AMP during the first few iterations, but they vanish quickly as $t$ approaches $T$, in return for over $50\%$ communication reduction beyond that provided by BT-MP-AMP, as shown in Table \ref{tb:Rtotal}. 

Note also that the ECSQ implementation of DP-MP-AMP has lower SDR's than that predicted by DP results based on the RD function at the beginning. This is because the $0.255$-bits gap only holds in the high rate limit. However, due to the robustness of SE to disturbances, and the increasingly high rates as $t$ approaches $T$, the ECSQ implementation matches the predicted DP results at the last iteration.

\section{Conclusion}
In this paper, we proposed a multi-processor approximate message passing framework with lossy compression. We used a uniform quantizer with entropy coding to reduce communication costs, and reformulated the state evolution formalism while accounting for quantization noise. Combining the quantizers and modified state evolution equation, an online back-tracking approach and another method based on dynamic programming determine the coding rate in each iteration by controlling the induced error. The numerical results suggest that our approaches can maintain a high signal-to-distortion-ratio despite a significant and often dramatic reduction in inter-processor communication costs.

\bibliographystyle{IEEEbib}
%\bibliography{strings,refs}
\bibliography{DAMP_ICASSP16}

\end{document}